\documentclass[11pt,a4paper]{article}

\usepackage{amsmath,amssymb,graphicx,booktabs}
\usepackage[margin=2.5cm]{geometry}
\usepackage{cite}

\title{Event-Level Voxel Reconstruction in Two-Photon Absorption Scans Using Pixel-Overlap Selection in Timepix3}

\author{
Tianqi Gao \\
\small Cavendish Laboratory, University of Cambridge
}

\date{}

\begin{document}
\maketitle

\begin{abstract}
Two-photon absorption (TPA) enables three-dimensional characterisation of silicon detectors by generating charge carriers within a confined volume around a focused laser spot. In combination with pixelated readout systems, TPA measurements provide access to spatially resolved timing observables relevant for electric field reconstruction. However, the interpretation of TPA data in segmented detectors is non-trivial: a single excitation produces multi-pixel clusters within the intrinsic time resolution of the readout, and in many implementations no external synchronisation between laser pulses and detector data is available.

In this work, we present a reconstruction framework for event-level voxelisation of TPA scans using Timepix3, operating on continuous, unsynchronised data. The method introduces a pixel-overlap-based definition of TPA events and a cluster-level timing estimator based on the highest deposited charge within a region of interest. This approach enables blind reconstruction of dwell structure and stable assignment of voxel timing without external triggers. We demonstrate that commonly used alternatives, such as centroid-based selection or earliest-hit timing, introduce systematic spatial biases in clustered events.

The proposed framework provides a robust and general method for reconstructing voxel-resolved timing information in segmented detectors, and is directly applicable to TPA-based studies of electric field distributions and charge transport in silicon sensors.
\end{abstract}

\section{Introduction}

The development of silicon detectors for high-energy physics and related applications increasingly requires detailed knowledge of charge transport and electric field distributions at the micrometre scale. Experimental techniques capable of resolving detector response in three dimensions are therefore of central importance. Among these, the Transient Current Technique (TCT) has established itself as a standard tool for studying charge collection and electric field properties in semiconductor devices.

The Two-Photon Absorption Transient Current Technique (TPA-TCT) extends conventional TCT by exploiting nonlinear optical absorption to generate charge carriers within a confined volume around the focal point of an infrared laser~\cite{Pape2023,Wiehe2021}. This enables true three-dimensional spatial resolution, allowing measurements of quantities such as charge collection, drift time, and electric field strength as a function of depth inside the detector. TPA-TCT has been widely applied to the characterisation of silicon sensors, including segmented devices relevant for modern detector systems~\cite{Pape2024}.

While the technique is well established for pad detectors and externally triggered systems, its application to pixelated readout introduces additional challenges. In segmented detectors, a single TPA excitation produces a cluster of activated pixels, all recorded within the intrinsic time resolution of the readout electronics~\cite{Timepix3}. As a result, the spatial origin of the excitation cannot be directly associated with a single pixel measurement. Furthermore, in asynchronous or triggerless readout systems such as Timepix3, the detector data are not synchronised to the laser pulses. The reconstruction of the temporal structure of the scan, including dwell assignment and event timing, must therefore be performed directly from the recorded data without an external reference.

These two effects — cluster formation and lack of synchronisation — lead to an ambiguity in associating measured signals with the underlying voxel of charge generation. Existing approaches often rely on centroid-based selections or earliest-hit timing estimators. However, these methods can introduce systematic biases, particularly in configurations where the excitation is not centred within the pixel or where charge sharing is significant.

In this work, we address these challenges by developing a reconstruction framework for event-level voxelisation of TPA scans in segmented detectors. The method is based on two key elements. First, TPA events are defined using a pixel-overlap criterion with respect to a region of interest, ensuring that all relevant cluster information is retained. Second, a cluster-level timing estimator is introduced, assigning the event time based on the pixel with the highest deposited charge within the region of interest. This choice is motivated by the physical localisation of charge generation and is shown to provide a stable and unbiased timing observable.

Importantly, the framework operates on continuous, unsynchronised data and does not require external triggering. The temporal structure of the scan, including dwell intervals and event grouping, is reconstructed directly from the data stream. This enables blind reconstruction of voxel-resolved timing information in systems where synchronisation is not available.

The method presented here provides a general approach for analysing TPA data in segmented detectors and forms the basis for subsequent measurements of electric field distributions and charge transport properties in silicon sensors.
\section{Experimental Overview}

Briefly describe:
\begin{itemize}
\item Timepix3-based readout
\item TPA laser parameters (wavelength, pulse duration)
\item scanning geometry (edge illumination, voxel stepping)
\item data structure (clustered events)
\end{itemize}

\section{Reconstruction Method}

This section describes the reconstruction framework used to extract voxel-resolved timing information from two-photon absorption (TPA) scans recorded with a pixelated detector. The method is designed to operate on continuous, unsynchronised data and addresses two key challenges: (i) the formation of multi-pixel clusters from a single excitation, and (ii) the absence of external timing reference between laser pulses and detector readout.

\subsection{Data Representation}

The detector data are recorded as a sequence of clustered events. Each event consists of a set of pixels $\{(x_i, y_i)\}$ with associated time-of-arrival (ToA) and time-over-threshold (ToT) values. A single TPA excitation produces a spatially extended cluster, reflecting the diffusion and drift of charge carriers within the sensor.

In the absence of external synchronisation, the data form a continuous time series of clustered events. The reconstruction must therefore identify both the spatial origin of each excitation and its temporal grouping into scan dwell intervals.

\subsection{Definition of Region of Interest}

A region of interest (ROI) is defined in pixel coordinates as:
\begin{equation}
(x, y) \in [x_{\mathrm{min}}, x_{\mathrm{max}}] \times [y_{\mathrm{min}}, y_{\mathrm{max}}]
\end{equation}

The ROI is chosen to encompass the expected location of the TPA excitation within the sensor. This region defines the spatial domain used for event selection and timing extraction.

\begin{figure}[htbp]
\centering
\includegraphics[width=0.1\textwidth]{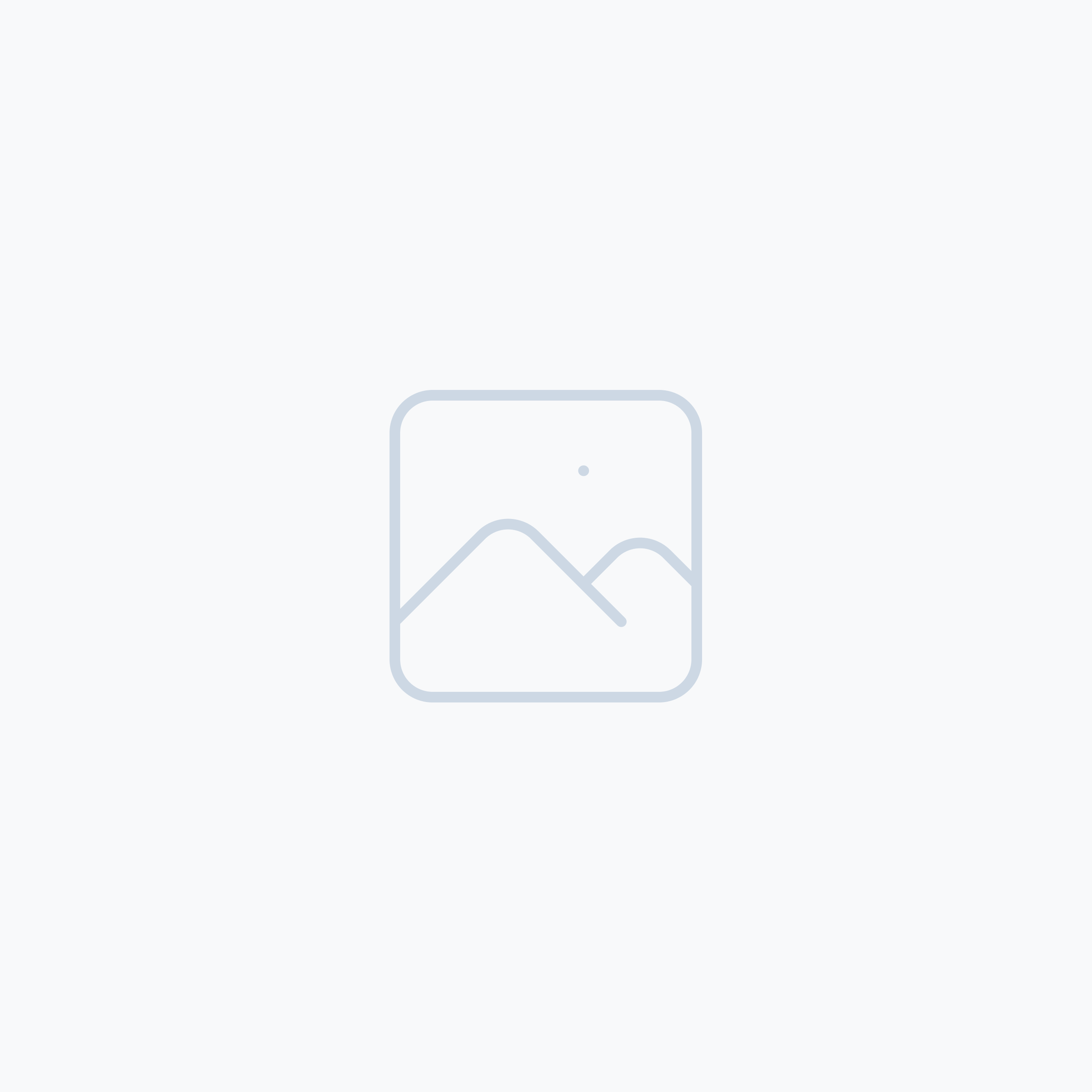}
\caption{Definition of the region of interest (ROI) within the pixel matrix. The ROI is selected to contain the expected TPA excitation region while excluding unrelated background activity.}
\label{fig:roi_definition}
\end{figure}

\subsection{Pixel-Overlap Event Selection}

TPA events are identified using a pixel-overlap criterion. A clustered event is classified as a TPA event if at least one of its constituent pixels lies within the ROI:
\begin{equation}
\exists\, i \;\text{s.t.}\; (x_i, y_i) \in \mathrm{ROI}
\end{equation}

\begin{figure}[htbp]
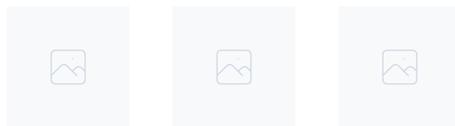

\centering

% Left: cluster + ROI overlap
\includegraphics[width=0.1\textwidth]{figures/placeholder.png}
\hspace{0.02\textwidth}
% Middle: centroid selection
\includegraphics[width=0.1\textwidth]{figures/placeholder.png}
\hspace{0.02\textwidth}
% Right: highest-ToT selection
\includegraphics[width=0.1\textwidth]{figures/placeholder.png}

\caption{
Conceptual illustration of reconstruction ambiguities in TPA measurements with pixelated detectors.
\textbf{Left:} A single TPA excitation produces a multi-pixel cluster, with only part of the cluster overlapping the region of interest (ROI).
\textbf{Centre:} Centroid-based selection may reject valid events or misrepresent the excitation position when the cluster is asymmetric or partially outside the ROI.
\textbf{Right:} Pixel-overlap selection retains all relevant events, while the highest-ToT pixel provides a stable estimator of the excitation location within the cluster.
}
\label{fig:conceptual_reconstruction}
\end{figure}

This selection retains the full cluster associated with each excitation while ensuring that the event is spatially relevant to the region under study.

This approach contrasts with centroid-based selection methods, where events are accepted only if the cluster centroid lies within the ROI. In the presence of asymmetric charge sharing or off-centre excitation, centroid selection can exclude relevant events or introduce spatial biases.

\begin{figure}[htbp]
\centering
\includegraphics[width=0.1\textwidth]{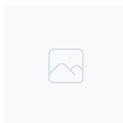}
\caption{Comparison of event selection strategies. Pixel-overlap selection accepts clusters that intersect the ROI, while centroid-based selection may reject such events if the cluster centroid lies outside the ROI. This leads to systematic biases in spatial reconstruction.}
\label{fig:pixel_overlap_vs_centroid}
\end{figure}

\subsection{Dwell Reconstruction from Continuous Data}

The temporal structure of the scan is reconstructed directly from the data stream. Events are grouped into dwell intervals based on their time separation.

Let $t_k$ denote the time of the $k$-th selected event. Consecutive events are assigned to the same dwell if:
\begin{equation}
t_{k+1} - t_k < \Delta t_{\mathrm{gap}}
\end{equation}

where $\Delta t_{\mathrm{gap}}$ is a threshold parameter defining the maximum allowed gap between events within a dwell.

This procedure partitions the event sequence into a set of dwell intervals:
\begin{equation}
\{ \mathcal{D}_j \}, \quad j = 1, 2, \dots
\end{equation}

Each dwell corresponds to a fixed laser position in the scan.

\begin{figure}[htbp]
\centering
\includegraphics[width=0.1\textwidth]{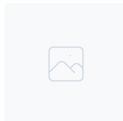}
\caption{Reconstruction of dwell intervals from continuous, unsynchronised data. Events are grouped based on their time separation, allowing the dwell structure of the scan to be recovered without external timing information.}
\label{fig:dwell_reconstruction}
\end{figure}

\subsection{Cluster-Level Timing Estimator}

For each TPA event, the timing must be assigned to a representative pixel within the cluster. Since all pixels in the cluster are recorded within the intrinsic detector time resolution, direct use of individual pixel ToA values is ambiguous.

We define a cluster-level timing estimator based on the pixel with the highest ToT within the ROI:
\begin{equation}
t_{\mathrm{event}} = \mathrm{ToA}_{i^*}, \quad i^* = \arg\max_{i \in \mathrm{ROI}} \mathrm{ToT}_i
\end{equation}

This choice is motivated by the localisation of charge generation: the pixel receiving the largest charge deposition is most likely to correspond to the position closest to the excitation point.

Alternative estimators, such as the earliest ToA within the cluster, are sensitive to edge effects and noise, and can lead to systematic spatial displacement of the reconstructed signal.

\begin{figure}[htbp]
\centering
\includegraphics[width=0.1\textwidth]{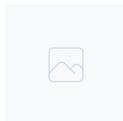}
\caption{Illustration of timing estimators within a clustered event. The earliest ToA pixel is often located at the periphery of the cluster, while the highest-ToT pixel is spatially closer to the centre of charge deposition and provides a more stable timing reference.}
\label{fig:timing_estimator}
\end{figure}

\subsection{Voxel Assignment}

Each reconstructed event is assigned to a voxel corresponding to its dwell index and spatial location within the ROI. The combination of dwell grouping and cluster-level timing provides a mapping:
\begin{equation}
(x, y, z) \rightarrow t_{\mathrm{event}}
\end{equation}

where the depth coordinate $z$ is determined by the scan position.

This enables the construction of voxel-resolved timing observables across the sensor volume.

\subsection{Summary of the Reconstruction Procedure}

The full reconstruction workflow is summarised as follows:

\begin{enumerate}
\item Select events using the pixel-overlap criterion with respect to the ROI
\item Group events into dwell intervals based on time separation
\item For each event, identify the highest-ToT pixel within the ROI
\item Assign event timing using the ToA of the selected pixel
\item Map events to spatial voxels based on dwell index and scan geometry
\end{enumerate}

\begin{figure}[htbp]
\centering
\includegraphics[width=0.1\textwidth]{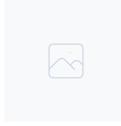}
\caption{Overview of the reconstruction workflow. Continuous detector data are filtered using pixel-overlap selection, grouped into dwell intervals, and processed with a cluster-level timing estimator to produce voxel-resolved timing information.}
\label{fig:workflow}
\end{figure}

\section{Validation of the Timing Estimator}

The choice of timing estimator within a clustered event is critical for accurate voxel reconstruction. In particular, the use of individual pixel ToA values is ambiguous due to the simultaneous activation of multiple pixels within the intrinsic time resolution of the detector. To validate the proposed highest-ToT timing estimator, we compare it with alternative definitions, most notably the earliest-ToA pixel within the region of interest.

For each TPA event, two candidate pixels are identified:
\begin{itemize}
\item the pixel with the highest ToT within the ROI
\item the pixel with the earliest ToA within the ROI
\end{itemize}

The spatial distributions of these two selections are compared.

\begin{figure}[htbp]
\centering
\includegraphics[width=0.1\textwidth]{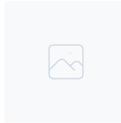}
\caption{Spatial distribution of pixels selected by the highest-ToT criterion. The distribution is localised and consistent with the expected region of charge deposition.}
\label{fig:highest_tot_map}
\end{figure}

\begin{figure}[htbp]
\centering
\includegraphics[width=0.1\textwidth]{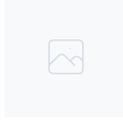}
\caption{Spatial distribution of pixels selected by the earliest-ToA criterion. The distribution is systematically shifted towards the cluster periphery, indicating sensitivity to edge effects.}
\label{fig:earliest_toa_map}
\end{figure}

To quantify the discrepancy between the two estimators, we define the spatial separation:
\begin{equation}
\Delta r = \sqrt{(x_{\mathrm{ToT}} - x_{\mathrm{ToA}})^2 + (y_{\mathrm{ToT}} - y_{\mathrm{ToA}})^2}
\end{equation}

\begin{figure}[htbp]
\centering
\includegraphics[width=0.1\textwidth]{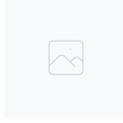}
\caption{Distribution of spatial separation between highest-ToT and earliest-ToA pixel selections. A broad distribution indicates that the two estimators frequently select different pixels within the cluster.}
\label{fig:delta_r}
\end{figure}

The results demonstrate that the earliest-ToA estimator is systematically displaced relative to the region of highest charge deposition. This behaviour is consistent with the propagation of charge to peripheral pixels and the sensitivity of ToA to low-threshold fluctuations. In contrast, the highest-ToT estimator provides a spatially stable and physically motivated representation of the excitation point.

\section{Definition of the TPA Region (TPAR)}

While the ROI defines a coarse spatial selection, the true region of charge generation is typically confined to a smaller subset of pixels. We define the TPA region (TPAR) as the set of pixels within the ROI that consistently receive charge from TPA events.

The TPAR is determined from accumulated cluster information across all selected events. Two complementary observables are used:
\begin{itemize}
\item cluster occupancy: frequency of pixel activation
\item highest-ToT pixel distribution: frequency of maximal charge deposition
\end{itemize}

\begin{figure}[htbp]
\centering
\includegraphics[width=0.1\textwidth]{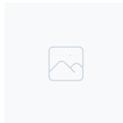}
\caption{Cluster occupancy map within the ROI, showing the spatial distribution of activated pixels across all TPA events. The high-occupancy region defines the approximate extent of the TPAR.}
\label{fig:cluster_occupancy}
\end{figure}

\begin{figure}[htbp]
\centering
\includegraphics[width=0.1\textwidth]{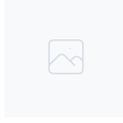}
\caption{Spatial distribution of highest-ToT pixels within the ROI. The concentration of events defines the core of the TPAR and provides a refined estimate of the excitation region.}
\label{fig:tpar_core}
\end{figure}

The TPAR is defined as the subset of ROI pixels satisfying a threshold on occupancy or highest-ToT frequency. This data-driven definition ensures that the selected region corresponds to the physical excitation volume, independent of assumptions about the scan geometry.

\section{General Applicability}

The reconstruction framework presented in this work is not limited to the specific detector configuration studied here. The method relies only on general properties of TPA measurements and segmented readout systems.

In particular, the pixel-overlap event definition is applicable to any detector where a localised excitation produces spatially extended clusters. The cluster-level timing estimator based on highest charge deposition provides a robust alternative to pixel-level timing in systems with limited temporal resolution.

The ability to reconstruct dwell structure from continuous, unsynchronised data is especially relevant for detectors operating in triggerless modes. This includes a broad class of modern pixel detectors used in high-rate environments.

Furthermore, the framework is directly compatible with downstream analyses, including voxel-resolved mapping of drift time, charge collection efficiency, and electric field distributions.

\section{Conclusion}

We have presented a reconstruction framework for voxel-level analysis of two-photon absorption scans in segmented silicon detectors. The method addresses two central challenges: the ambiguity introduced by cluster formation and the absence of external synchronisation between laser and detector.

By defining TPA events through pixel-overlap selection and assigning timing based on the highest-ToT pixel within the ROI, we obtain a stable and physically meaningful estimator of the excitation point. The temporal structure of the scan is reconstructed directly from continuous data, enabling blind identification of dwell intervals.

Validation studies show that alternative estimators, such as earliest-ToA selection, introduce systematic spatial biases. The proposed approach avoids these effects and provides a consistent basis for voxel-resolved measurements.

The framework establishes a general methodology for analysing TPA data in pixelated detectors and provides the foundation for precise studies of charge transport and electric field distributions in silicon sensors.

\bibliographystyle{unsrt}
\bibliography{refs}

\end{document}